\documentclass[12pt]{article}
\usepackage{epsf}
\usepackage{amsmath}
\usepackage{graphics}
\usepackage{cite}

\renewcommand{\thefootnote}{\#\arabic{footnote}}
\begin{document}

\newcommand{\gtrsim}{ \mathop{}_{\textstyle \sim}^{\textstyle >} }
\newcommand{\lesssim}{ \mathop{}_{\textstyle \sim}^{\textstyle <} }

\newcommand{\rem}[1]{{\bf #1}}

\renewcommand{\thefootnote}{\fnsymbol{footnote}}
\setcounter{footnote}{0}
\begin{titlepage}

\def\thefootnote{\fnsymbol{footnote}}

\begin{center}

\vskip .5in
\bigskip
\bigskip
{\Large \bf Past Eras In Cyclic Cosmological Models} 

\vskip .45in

{\bf Paul H. Frampton\footnote{frampton@physics.unc.edu}}

\bigskip
\bigskip

{Department of Physics and Astronomy, UNC-Chapel Hill, NC 27599.}

\end{center}

\vskip .4in

\begin{abstract}
In infinitely cyclic cosmology past eras
are discussed using set theory
and transfinite numbers. One 
consistent scenario, already in the
literature, is where there is always
a countably infinite number, $\aleph_0$, 
of universes and no big bang. 
I describe here an alternative 
where the present number of universes is
$\aleph_0$ and in the infinite past
there was only a finite number of universes. 
In this alternative model it is 
also possible that 
there was no big bang.
\end{abstract}

\end{titlepage}

\renewcommand{\thepage}{\arabic{page}}
\setcounter{page}{1}
\renewcommand{\thefootnote}{\#\arabic{footnote}}

\newpage

\noindent {\it Introduction.}

\bigskip
\bigskip

\noindent It was suggested already in the first half
of the twentieth century, most emphatically by Tolman
\cite{Tolman}, that the initial singularity of
a big bang cosmology might be avoided by
hypothesizing a cyclic universe in which the expansion
era ends at a turnaround to a contraction era which itself
ends in a bounce.

\bigskip

\noindent Implementation of cyclic cosmology
confronted, however, a seemingly impossible problem which 
we shall call the {\it Tolman conundrum}. 
The Tolman conundrum is that, because
entropy monotonically increases, future 
cycles are larger and longer while
in the past cycles were smaller and
shorter leading at a finite past time
to a big bang. The motivation of
avoiding a big bang was thus frustrated.

\bigskip

\noindent It is fair to say that the Tolman conundrum
and the failure to solve it led
to the continued acceptance of the big bang
from Tolman's time until the twenty-first
century. Incidentally it is worth remarking 
that in his work Tolman never 
found it necessary to assume the
possibility of more than one universe.

\bigskip

\noindent A major observational discovery in cosmology was of
the accelerated expansion rate of the universe
and the concomitant dark energy. The idea that
dark energy might aid in constructing a consistent
cyclic universe was pioneered in a useful
and important series of papers by Steinhardt
and Turok\cite{SteinhardtTurok}. These authors
conceived of the idea that branes colliding
in an extra space dimension could provide an
alternative to inflationary cosmology and,
beyond that, underly a cyclic cosmology.
These papers contain a large number of 
important new ideas
and make considerable progress towards avoiding
a big bang. They do not, however, solve 
the Tolman conundrum.

\bigskip

\noindent In a different approach
\footnote{Other recent works on cyclic models are listed in \cite{catchall}}
Baum and the author \cite{BF}
solved the Tolman conundrum while disregarding
other obstacles in the hope that they could be solved 
if the Tolman conundrum is.

\bigskip

\noindent The resultant model leads to
some unexpected but logically inevitable
assumptions needed to reduce the
dimensionless entropy of the universe from
over one googol ($10^{100}$) to
zero at turnaround. 

\bigskip

\noindent These are those assumptions:

\bigskip

\noindent (i) The dark energy must have 
equation of state $w = p/\rho < -1$, with
$|(w+1)|$ arbitrarily small. This ensures approach
to a big rip and possible fragmentation at turnaround
into more than one googol of causally
disconnected patches.

\bigskip

\noindent (ii) The universe contracts adiabatically
with zero entropy, empty of
matter and containing only dark energy. This avoids
the otherwise necessary fine tuning of initial
conditions at the beginning of the expansion era
to less than a part in one googolplex ($10^{10^{100}}$).

\bigskip

\noindent In the model of \cite{SteinhardtTurok}
a difficult technical problem concerns the singularity
at the bounce.
In the complementary model of \cite{BF} it is the turnaround
that requires as an equally challenging problem
the computation of entropy in
extreme spacetime backgrounds.

\bigskip
\bigskip
\bigskip
\bigskip

\noindent {\it The number of universes}

\bigskip
\bigskip

\noindent In the model of \cite{BF} 
the following relationship between the number
of universes $\Sigma_n$ after cycle $n$ and $\Sigma_{n+1}$
after cycle $(n+1)$
occurs
\begin{equation}
\Sigma_{n+1} = N \Sigma_n
\label{Nequation}
\end{equation}

\noindent where $N$ is a finite number necessarily
bigger than one googol. One may take {\it e.g.}
$N=10^{123}$ as the number of universes spawned
at each turnaround.

\bigskip

\noindent Let me label the present expansion era
as $n = 0$. It is straightforward to see that to avoid a big bang
the present number $\Sigma_0$ must be infinite.
I therefore set it equal to the smallest infinite ordinal
$\Sigma_0 = \aleph_0$.

\bigskip

\noindent Higher transfinites than $\aleph_0$ while possible to define precisely
are more difficult to describe \cite{Cantor,Hausdorff,Woodin}.
They are surely irrelevant to cosmology.

\bigskip

\noindent With these preliminaries I can now define
two specific cyclic models which I shall represent
by exhibiting time-ordered sets $S_1$ and $S_2$
of their respective transfinite sequences
of cardinal $\Sigma_n$.

\bigskip
\bigskip
\bigskip

\noindent {\it The first model}

\bigskip
\bigskip

\noindent 
$S_1$ will characterize my first model and $S_2$ my second.

\bigskip

\noindent I first note that multiplication of $\aleph_0$ by any finite number
leaves it unchanged as $\aleph_0$, for example

\begin{equation}
N^p \aleph_0 \equiv \aleph_0
\label{aleph}
\end{equation}
for all finite $p$.

\bigskip
\bigskip

\noindent I introduce the notation that
a double ellipsis .. denotes a finite number of elements
while a triple ellipsis ... denotes a countably infinite 
number ($\aleph_0$) of elements.

\bigskip
\bigskip

\noindent I am now ready to define the first cyclic model by the 
transfinite sequence $S_1$. 

\begin{equation}
S_1 = \left[ ..., \aleph_0, \aleph_0, \aleph_0, ... \right]
\label{S1}
\end{equation}

\bigskip

\noindent This implies that the model always contains
an infinite number $\aleph_0$ of universes\footnote{Because entropy
is constant at $\aleph_0$ I can say that
this cyclic model eliminates an arrow of time.}.
It
was the only model discussed in the second paper of \cite{BF}.
The transfinite sequence $S_1$ is equivalent to it own
anti-time-ordering $S_1^*$ and I can therefore write: $S_1 \equiv S_1^*$.

\bigskip
\bigskip

\noindent At first sight, $S_1$ seems the only consistent possibilty.
However, I believe there exists an alternative cyclic model characterized
by a strictly time-ordered transfinite sequence $S_2$.

\bigskip
\bigskip
\bigskip

\noindent {\it The alternative model}

\bigskip
\bigskip

\noindent 

\noindent Can a cyclic model exist which begins with a finite number
of universes, for example one universe?

\bigskip

\noindent I have reached a positive conclusion
by considering a time-ordered transfinite sequence
of cardinals $S_2$ with $S_2 \not\equiv S_2^*$ as follows

\begin{equation}
S_2 = [1, N, N^2, .. ,N^p, N^{p+1}, N^{p+2}, ..., \aleph_0,
\aleph_0, \aleph_0, ...]
\label{S2}
\end{equation}
where $p$, like $N$, is finite and we employ Eq.(\ref{aleph}).

\bigskip
\bigskip

\noindent In the transfinite sequence$S_2$ 
the first infinite element $\aleph_0$
must have absence of a precedent. We hypothesize that $S_2$ 
corresponds to a cyclic cosmology.
Provided that at the present time $t=t_0$
there is an infinite number $\aleph_0$ of universes
there was no big bang.

\bigskip
\bigskip

\noindent This introduces an interesting concept from
mathematics because while  physically in the model\cite{BF}
it is puzzling if 
a finite number of universes can spawn an infinite number
yet this is expressed mathematically by the useful
notion {\it absence of precedent} which was
unnecessary in $S_1$.

\bigskip
\bigskip

\noindent This is just as in set theory the first transfinite
ordinal coming after 1, 2, 3, ... is $\aleph_0$
which has a following transfinite $\aleph_0 + 1$
but $\aleph_0$ has no precedent.

\bigskip
\bigskip

\noindent All that is necessary to eliminate a big bang
in Eq.(\ref{S2}) is that the present universe correspond to a cardinal
of $S_2$ preceded by a triple ellipsis.

\bigskip
\bigskip

\noindent These considerations are applicable
to all cyclic models\cite{SteinhardtTurok,BF,catchall}
which solve the Tolman conundrum.

\newpage

\noindent {\it Discussion}

\bigskip
\bigskip

\noindent The distinction between the first and the alternate
model is that in the alternate model we both avoid a big bang
and allow the original creation to be of a finite number
of universes possibly only one universe.
One can say that it is easier to create only one universe 
as in the alternate model
than go to the bother of creating an infinite number
$\aleph_0$ of universes as in the first model.

\bigskip
\bigskip

\noindent Both models are characterized by a present
equation of state of dark energy satisfying
$w < -1$ as will be tested by the Planck mission
and subsequent observations.

\bigskip

\newpage

\begin{center}

{\bf Acknowledgement}

\end{center}

\bigskip

\noindent 
This work was supported in part by 
the U.S. Department of Energy under Grant number DE-FG02-06ER41418.

\end{document}